\documentclass[referee]{aa} 

\usepackage{graphicx}
\usepackage{txfonts}
 \usepackage{mathtools}
\begin{document}

   \title{ Chemical analysis of  very metal-poor turn-off stars from SDSS - DR12   }
%
\titlerunning{ EMP stars}
   \author{P. Fran\c cois
          \inst{1,2}
          \and 
          E. Caffau\inst{3}
          \and 
          S. Wanajo \inst{4, 5}
          \and 
          D. Aguado \inst{6,7}
          \and
          M. Spite \inst{3}
          \and
           M. Aoki \inst{8}
           \and
            W. Aoki \inst{9}
            \and
             P. Bonifacio \inst{3}
             \and 
             A. J.  Gallagher \inst{10,3}
             \and 
              S. Salvadori \inst{3,11,12}
             \and  
             F. Spite   \inst{3}
            \thanks{Based on observations collected at the European Organisation for Astronomical Research in the Southern Hemisphere under  ESO programme ID 099.D-0576(A).}
          }
   \institute{GEPI, Observatoire de Paris, PSL Research University, CNRS, 61 Avenue de l'Observatoire, 75014 Paris, France \\
              \email{patrick.francois@obspm.fr}
         \and
             UPJV, Universit\'e de Picardie Jules Verne, 33 rue St Leu, 80080 Amiens, France 
          \and
          GEPI, Observatoire de Paris, PSL Research University, CNRS, Place Jules Janssen, 92190 Meudon, France    
          \and
          Department of Engineering and Applied Sciences, Sophia University, Chiyoda-ku, Tokyo 102-8554, Japan
          \and
          iTHEMS Research Group, RIKEN, Wako, Saitama 351-0198, Japan
          \and
          Instituto de Astrof\`{i}sica de Canarias, V\'{i}a L\'{a}ctea, E-38205, La Laguna, Tenerife, Spain Canarias
          \and
          Universidad de La Laguna, Departamento de Astrof\'isica, 38206 La Laguna, Tenerife, Spain
          \and
          European Southern Observatory, Karl-Schwarzschild-Str. 2   85748 Garching bei Muenchen, Germany
          \and
          National Observatory of Japan, Mitaka, Tokyo, Japan
          \and
           Max-Planck-Institut f\"{u}r Astronomie, K\"{o}nigstuhl 17, 69117 Heidelberg, Germany   
           \and
           Dipartimento di Fisica e Astronomia, Universit\'{a} di Firenze, Via G. Sansone 1, Sesto Fiorentino, Italy
           \and
           INAF/Osservatorio Astrofisico di Arcetri, Largo E. Fermi 5, Firenze, Italy
             }
   \date{Received  11 July 2018  / Accepted  13 August 2018  }

 
  \abstract
   {The most metal-poor stars  are the relics of the early chemical evolution of the Galaxy.  Their chemical composition is an important tool to  constrain the nucleosynthesis in the first generation of stars.   
   The aim is to observe a sample of  extremely metal-poor star (EMP stars)  candidates selected from the Sloan Digital Sky  Survey Data Release 12  (SDSS DR12) and  determine their chemical composition.}
   {We obtain medium resolution spectra of  a sample of six stars using   the X-Shooter spectrograph at the Very Large Telescope (VLT)  and we used  ATLAS  models to compute the  abundances.     }
   {Five  stars of the sample have a metallicity [Fe/H] between  -2.5~dex and -3.2~dex. We confirm the recent discovery of  SDSS~J002314.00+030758.0.   As a star with an extremely low [Fe/H] ratio. Assuming the $\alpha$-enhancement  [Ca/Fe] = +0.4 dex, we obtain [Fe/H] = -6.1 dex.  }
{    We could also determine its magnesium abundance and found that this star exhibits a very high ratio [Mg/Fe] $\ge$ +3.60~dex assuming [Fe/H] = -6.13 dex. We determined the carbon abundance and found A(C) = 6.4 dex. From this carbon abundance, this stars belongs to the lower band of the A(C) - [Fe/H] diagram.
   }
   {}
   
                \authorrunning{P. Fran\c{c}ois et al.}
\titlerunning{ EMP stars}               
   \maketitle
%

\section{Introduction}

 The lambda Cold Dark Matter ( $\Lambda$-CDM) cosmological model has  received an impressive confirmation from the Wilkinson Microwave Anisotropy Probe (WMAP)  and PLANCK  satellites \citep[and references therein]{planck2016} over in recent years. The average redshift at which reionization occurs is found to lie between z = 7.8 and 8.8, depending on the model of reionization adopted. It may be that all the first stars were massive or exceedingly massive, with a very short lifetime \citep{bromm2009}, although other more recent  numerical simulations suggest that the distribution of possible masses of the first stars may be much broader than previously believed \citep{hirano2015}, and may even extend down to solar mass or below \citep{greif2011} leading to stars which are still alive and observable today. Until recently, the deepest survey searching for metal-poor stars was the Hamburg-ESO survey (HES), which reached V=16  \citep{christlieb2008}, although the first aim of this survey  was the search for distant quasars.
Thanks to the Hamburg-ESO survey,  several stars of extremely low iron content have been discovered. However, these stars are extremely rich in C and O, so that their overall metal content is  in fact comparable to the metal  content of Globular Cluster stars, with [M/H]  $\simeq$-2.3 dex. 

 Major progress with respect to HES  can be found  in an exploration of  the data of the Sloan Digital Sky Survey (SDSS). \citet{ludwig2008}  have developed an analysis tool that allows us to estimate the metallicity of Turn-off (TO) stars from the low resolution SDSS spectra. This tool can be used to derive the metallicity  and to select extremely metal-poor candidates using the strongest lines (Calcium H\&K) in the spectrum  used as a proxy for the metallicity.
It is, however, not yet possible to firmly evaluate   the metallicity  precisely below [Fe/H] = -3.0 dex,  because at  very low metallicity  the metallic lines become almost impossible to detect  at the resolution of the SDSS spectra \citep{aoki2013}. 
 The only solution to alleviate this degeneracy is  to observe these candidates at a higher resolution  to confirm their metallicities. The SDSS-based metallicities found by the method of \citet{ludwig2008} are essentially confirmed by the analysis at higher resolution. 
 This method has been  very successful and has led to a series of papers that have already been published  \citep{caffau2011a, caffau2011b, caffau2012, caffau2014, bonifacio2015, caffau2016, bonifacio2018}. 
 The most metal-poor stars  are formed out of gas that has been very likely enriched  by the ejecta of a single or a few supernovae.  From the determination of the chemical composition of these stars, we can derive important constraints on the nucleosynthesis in the first generation of stars who enrich the primordial gas and on the chemical inhomogeneites during the early evolution of our Galaxy. 
 These abundance determinations can also be used to constrain the scenario of formation of the first  low mass stars \citep{caffau2011a}.
 
 From the recent analysis of  SDSS DR12 data, we have detected new extremely metal-poor  candidates that have never been observed at high resolution. In this article, we report  the detailed analysis of six new extremely  metal-poor candidates observed with the X-Shooter  spectrograph installed on  Kueyen  at the ESO Very Large Telescope (VLT) on Cerro Paranal in Chile.  Similar observations have been conducted for a different set of stars  from  the northern hemisphere  at the Subaru telescope using  the High Dispersion spectrograph (HDS) in the framework of a French-Japanese collaboration.

\section{Observations}
The observations were performed in service mode with Kueyen
(VLT Unit 2 ) and the high-efficiency spectrograph X-Shooter   
\citep{dodo2006, vernet2011}.  The log book of the observations is reported in Table \ref{obslog}. 
The X-Shooter spectra
range from 300~nm to 2400~nm and are registered by three detectors.
The observations have been performed in staring mode
with 1 $\times$ 1 binning and a slit width of 0.8 \arcsec for the ultra-violet (UVB) arm and and  0.9 \arcsec
for the visible (VIS) arm.  
This corresponds to a resolving power of R = 6190 in
the UVB arm and R = 7410 in the VIS arm. The stellar light is divided between the three arms by X-Shooter.
We analysed here only the UVB and VIS spectra. The stars we
observed are faint and have most of their flux in the blue part
of the spectrum, so that the signal-to-noise ratio (S/N) of the
infra-red spectra is too low to allow the analysis. The spectra
were reduced using the X-Shooter pipeline \citep{goldo2006},
which performs the bias and background subtraction, cosmic ray-hit removal \citep{vandok2001}, sky subtraction \citep{kelson2003}
, flat-fielding, order extraction, and merging. 

%
   \begin{figure*}
   \centering
   \includegraphics[width=10cm, angle=0]{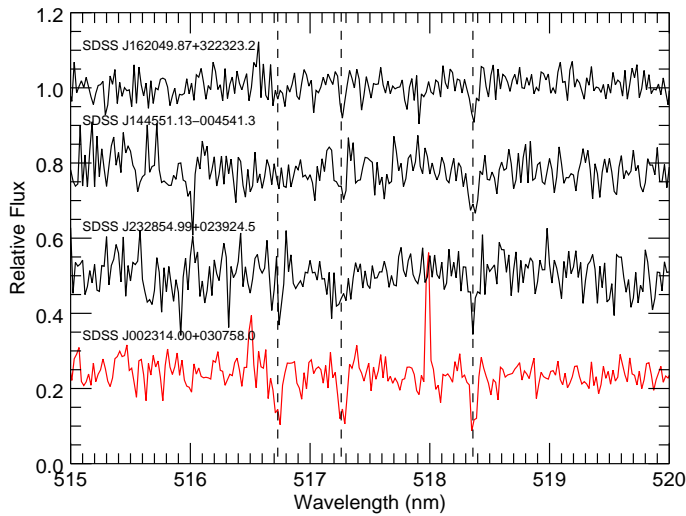}
      \caption{ X-Shooter spectra  of the stars centred  on the magnesium triplet. The continuum  of the three spectra located in the lower part of the plot has been shifted downward for clarity.  The vertical lines indicate the location of the magnesium absorption lines.}
              
         \label{spec}
   \end{figure*}

Figure \ref{spec} shows the spectra of some stars of the sample centred on the magnesium triplet. 
The spectra shown on this figure  correspond to the stars for which magnesium abundances have been determined.


\begin{table*}
 \caption{Observation log. } 
\label{obslog}
\centering
\begin{tabular}{l c c c c }
\hline\hline
  Object          &  g magnitude  &  Observation date &  Exp. time  [s]  &S/N @ 450 nm  \\      
\hline
        SDSS~J232854.99$+$023924.5      & 18.84 &       2017-06-03T08:10    & 3300.   &  25 \\  
        SDSS~J113207.12$-$082657.3      & 18.16 &       2017-06-21T00:16         & 1800. & 20 \\  
        SDSS~J002314.00$+$030758.0      & 17.91 &       2017-06-30T09:34         & 1800. & 31 \\  
        SDSS~J144551.13$-$004541.3      & 18.85 &       2017-07-16T00:49         &  3300.  &  25 \\  
        SDSS~J003730.31$+$245750.6      & 18.67 &  2017-08-23T07:31     &  2700.& 20 \\  
        SDSS~J003730.31$+$245750.6      &  18.67 &      2017-07-22T07:21         & 2700.& 15 \\  
        SDSS~J162049.87$+$322323.2      &  18.00 &      2017-07-22T00:17         & 1800. & 35  \\  
        SDSS~J002314.00$+$030758.0      & 17.91  &      2017-07-21T08:56     &   1800.  & 25 \\   

  \hline
  \end{tabular}
   \end{table*}

\section{Stellar parameters}

%
   \begin{figure*}
   \centering
   \includegraphics[width=10cm, angle=0]{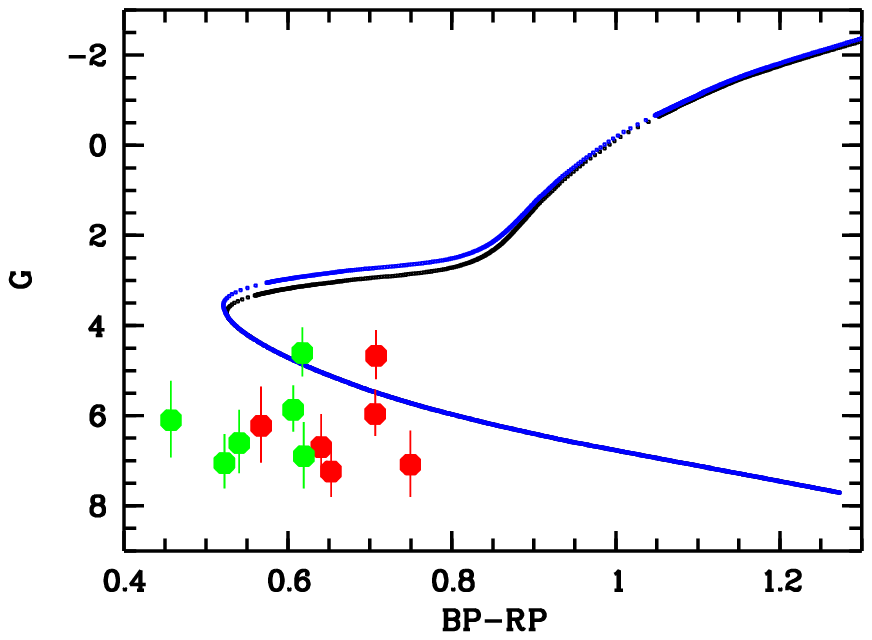}
      \caption{ Comparison of the location of the stars and isochrones on a G magnitude versus (BP-RP) magnitude diagram. The two  isochrones on the plot have been computed for an age of 14 Gyr (black line : $z=2.10^{-5}$ and blue line : $z=2.10^{-6}$).   BP-RP (BP and RP are the magnitudes measured respectively by the two  low resolution spectrographs, the Blue Photometer (BP) and the Red Photometer (RP)  onboard the Gaia satellite) versus the absolute G magnitude, with two different hypotheses for the reddening.  
The red dots represent the stars assuming  E(BP-RP)=0  whereas the green dots represent our stars' photometry with an extinction  correction E(BP-RP) using the E(B-V) correction  of PanSTARRS \citep{chambers2016}    and the conversion relations  E(BP-RP) = E(B-V)+0.07 and $A_{G}$=$A_{V}$=3.1  $\times$ E(B-V) (where $A_{G}$ and $A_{V}$ are the extinction in the G and the V band)  from \citet{andrae2018}.
              }
         \label{gaia_cmd}
   \end{figure*}

The stellar parameters have been derived  taking into account the SDSS photometry.
The effective temperatures in Table \ref{stellar_parameters} have  been computed by \citet{caffau2013a} . 
The effective temperature has been derived from the photometry, using the 
$(g-z)_{0}$ colour and the calibration described in \citet{ludwig2008} taking into account  the 
reddening according to the \citet{schlegel1998} extinction maps and corrected
as in \citet{bonifacio2000}. 
We also determine the spectroscopic temperatures by fitting the $H_{\alpha}$ line.  For the first four stars of Table \ref{stellar_parameters}, we found slightly  lower temperatures  ($\simeq 200 K$) than the photometric temperatures.
For the two remaining stars, SDSS~J003730.31+245750.6 and SDSS~J162049.87+322323.2, it was  difficult to obtain  a reliable estimate of the temperature due to the low S/N ratio of the spectra.  


The Gaia parallaxes  \citep[Gaia Collaboration et al. 2018]{arenou2018} for our stars are very imprecise; one star has a negative
parallax and the others all have a relative error $\sim 50$\% or larger
(up to 2100\%).
In order to get some insight into the luminosities of our stars,
we used the distance estimates of \citet{BJ18}.

\citet{BJ18} use a probabilistic approach to distance estimation. 
The first hypothesis is that the Gaia parallaxes have a Gaussian 
likelihood, and for each star the mean value of the distribution
and its standard deviation are given in the Gaia catalogue
in the columns {\tt parallax} and {\tt parallax\_error,} respectively. 
To this they add a prior of an exponentially 
decreasing space density\footnote{
\begin{equation}
P({\ensuremath{r}}{\ensuremath{\hspace{0.05em}\mid\hspace{0.05em}}}{\ensuremath{L}})  \ = \  \begin{dcases}
  \ \frac{1}{2{\ensuremath{L}}^3}\,{\ensuremath{r}}^2e^{-{\ensuremath{r}}/{\ensuremath{L}}}  & \:{\rm if}~~ {\ensuremath{r}} >0 \\
  \ 0                          & \:{\rm otherwise}
\end{dcases}
.\end{equation}
},
where $L > 0$ is a length scale \citep[see also][]{Bailer2015,Astraatmadja2016}.
In the model of \citet{BJ18}, $L$
varies with Galactic longitude and latitude, $(l,b)$. This amounts to  making
an assumption on the spatial structure of the Galaxy. 
In this sense these distances are biased and should not be over-interpreted. 
We nevertheless proceed with caution to use these distance estimates 
to see if they can help us discriminate between dwarfs and sub-giants.
Combining the Gaussian likelihood and the prior \citet{BJ18}
obtain a posterior probability distribution for the distance
to each star. They use the mode of this distribution as their 
distance estimate and they define two distances, $r_{lo}$
and $r_{hi}$ , such that the probability that the true distance
$r$ lies in the interval $[r_{lo},r_{hi}]$ is 0.6827.
In this sense this interval is akin to a $\pm 1 \sigma$
interval of a Gaussian distribution. However, 
since the posterior probability defined by \citet{BJ18}
is asymmetric, the estimated distance is not at the centre
of this interval. We use 
the estimated distance to compute the absolute magnitude for
each star and
$r_{lo}$ and $r_{hi}$ as estimates of the 
error on the distance and translate these to errors in 
estimated absolute magnitudes.

On Figure \ref{gaia_cmd}, we plotted  BP-RP  versus the absolute G magnitude with two different hypotheses for the reddening (BP and RP are the magnitudes measured respectively by the two  low resolution spectrographs, the Blue Photometer (BP) and the Red Photometer (RP)  on-board the Gaia satellite).  
The red dots represent the stars assuming 
E(BP-RP)=0  whereas the green dots represent our stars' photometry with an extinction  correction E(BP-RP) using the E(B-V) correction  of the The Panoramic Survey Telescope and Rapid Response System (PanSTARRS) \citep{chambers2016}    and the conversion relations  E(BP-RP) = E(B-V)+0.07 and $A_{G}$=$A_{V}$=3.1  $\times$ E(B-V) (where $A_{G}$ and $A_{V}$ are the extinction in the G and the V band)  from \citet{andrae2018} . We could then compare the location of our stars with the isochrones  computed by Chieffi (private communication). The age has been set at t=14 Gyrs and the two metallicities are   $z=2.10^{-5}$ and $z=2.10^{-6}$.

It is clear from Fig. \ref{gaia_cmd} that all our stars are warmer
than even the lowest luminosity giant stars, whatever the assumption
on the reddening. We can thus safely exclude the possibility that any of them is an evolved 
giant. An ambiguity exists, however, between stars that are
on the main sequence (MS) and sub-giant stars.
Considering the isochrones shown in Fig. \ref{gaia_cmd}, over the colour range spanned by
our stars, the mean log g of MS stars is 4.5 and that of sub-giant
stars 3.7 (for $Z=10^{-6}$) or 3.8 (for $Z=10^{-5}$).
The turn-off corresponds to log = 4.1.
Noting that, with the distance estimates of \citet{BJ18} none of the stars
is luminous enough to be considered a sub-giant, we make the simplifying
assumption that they are all MS and assume log g = 4.5.
This assumption has no impact on the derived abundances of the neutral species,
it affects only the singly  ionized species.

We attribute the fact that the stars do not fall neatly on any of the isochrones
to the limitation of the distance estimates of \citet{BJ18}. We consider
it likely that all the stars are MS, but we cannot robustly exclude the possibility that 
any of them is a TO or an early sub-giant. 
To understand what would be the implication of this,
we derived abundances also assuming a lower gravity of log g = 4.0. 
Taken at face value, the luminosities derived from the distance estimates
of \citet{BJ18} would imply higher surface gravities such as log g = 5.0
or even 5.5. However, since no isochrone predicts such high gravities
for warm stars, like those in the present sample, we discard this possibility.

 \begin{table*}
 \caption{Adopted stellar parameters for the list of targets. } 
\label{stellar_parameters}
\centering
\begin{tabular}{l c c c }
\hline\hline
  Object          &  $T_{eff}$ & log~g  & [Fe/H]    \\      
\hline
        SDSS~J232854.99+023924.5  &     631     0   &4.5    &  -3.0 \\  
        SDSS~J113207.12-082657.3        &       6420    &4.5 & -3.0 \\  
        SDSS~J002314.00+030758.0        &       6160    &4.5 & -5.0 \\  
        SDSS~J144551.13-004541.3                &       6530    & 4.5  &  -3.0 \\  
        SDSS~J003730.31+245750.6        &       6350    &  4.5 & -3.0 \\  
        SDSS~J162049.87+322323.2        &       6050    & 4.5  & -3.5  \\  
  \hline
  \end{tabular}
   \end{table*}
   
  \section{Analysis}  
 
 The determination of the abundances has been done  line by line using a grid of synthetic spectra, computed with the local thermodynamical equilibrium (LTE) spectral synthesis code  turbospectrum \citep{alvarez1998, plez2012} and based on 
 ATLAS models. From a set of well-selected  absorption lines, the code computes the abundance using fitting techniques. The best-fit profile is obtained by interpolating  in a series of pre-calculated spectra. The synthetic spectra have been computed with ATLAS9 models. However, for SDSS~J0023+0307,  which has a very peculiar chemical composition, we relied on  ATLAS12 models that allow us to compute models with a non-standard  chemical composition mix.  The procedure to derive the  chemical composition is described by \citet{caffau2013b} and has been used in several other papers  on the determination of the chemical composition of  extremely metal-poor stars \citep{bonifacio2015, caffau2016}. We adopted the solar abundances of \citet{caffau2011b} for C and Fe and the abundances of \citet{lodders09} for the other elements.

\section{Errors}

\begin{table}
 \caption{Estimated errors in the element abundance ratios [X/Fe] for the star  SDSS~J232854.99+023924.5. The other stars give similar results. } 
\label{errors}
\centering
\begin{tabular}{l r r r }
\hline\hline
  [X/Fe]            &   $\Delta T_{eff}$ =   & $\Delta$ log~g =   &$\Delta$  $ v_{t}$ =   \\     
                        &       100~K                 &    0.5~dex        &     0.5~km/s \\  
\hline
C    &   0.2       &   0.2      &     0.1    \\
Mg &   0.1  &  0.15 &  0.15  \\
Ca &   0.1  &  0.1  &  0.15 \\  
Si &   0.1 &  0.15  &  0.15  \\
Sr &   0.1  &  0.2  &  0.25  \\
Ba &   0.1  &  0.2   &  0.3  \\
  \hline
  \end{tabular}
   \end{table}

Table \ref{errors} lists the computed errors in the elemental abundances ratios due to typical uncertainties in the stellar parameters. The errors were estimated varying  $T_{eff}$ by $\pm$ 100~K, log~g  by $\pm$  0.5~dex, and $ v_{t}$  by $\pm$ 0.5 dex in the model atmosphere of SDSS~J232854.99$+$023924.5; other stars give similar results. In this star, we could measure the Mg, Si, Ca , Ba, and Sr abundances.  The main uncertainty comes from the error in the placement of the continuum when the synthetic line profiles are matched to the observed spectra. This error is of the order of 0.2 to 0.5 depending on the species under consideration, the largest value being for the neutron capture elements.  When several lines are available, the typical line to line scatter for a given elements is 0.1 to 0.2 dex.  We can also see from Table \ref{errors} that assuming a lower gravity for the stars (i.e. log g = 4.0 instead of log g=4.5 dex) shifts the abundances by at most 0.2 dex.

 \section{Results and discussion}
 
The abundance results  for the five less metal-poor stars have been gathered in Tables \ref{Fe_abundance} and \ref{abundances}, and the results for SDSS~J002314.00$+$030758.0 are reported in the next section. 

    \begin{figure*}
   \centering
   \includegraphics[width=15cm]{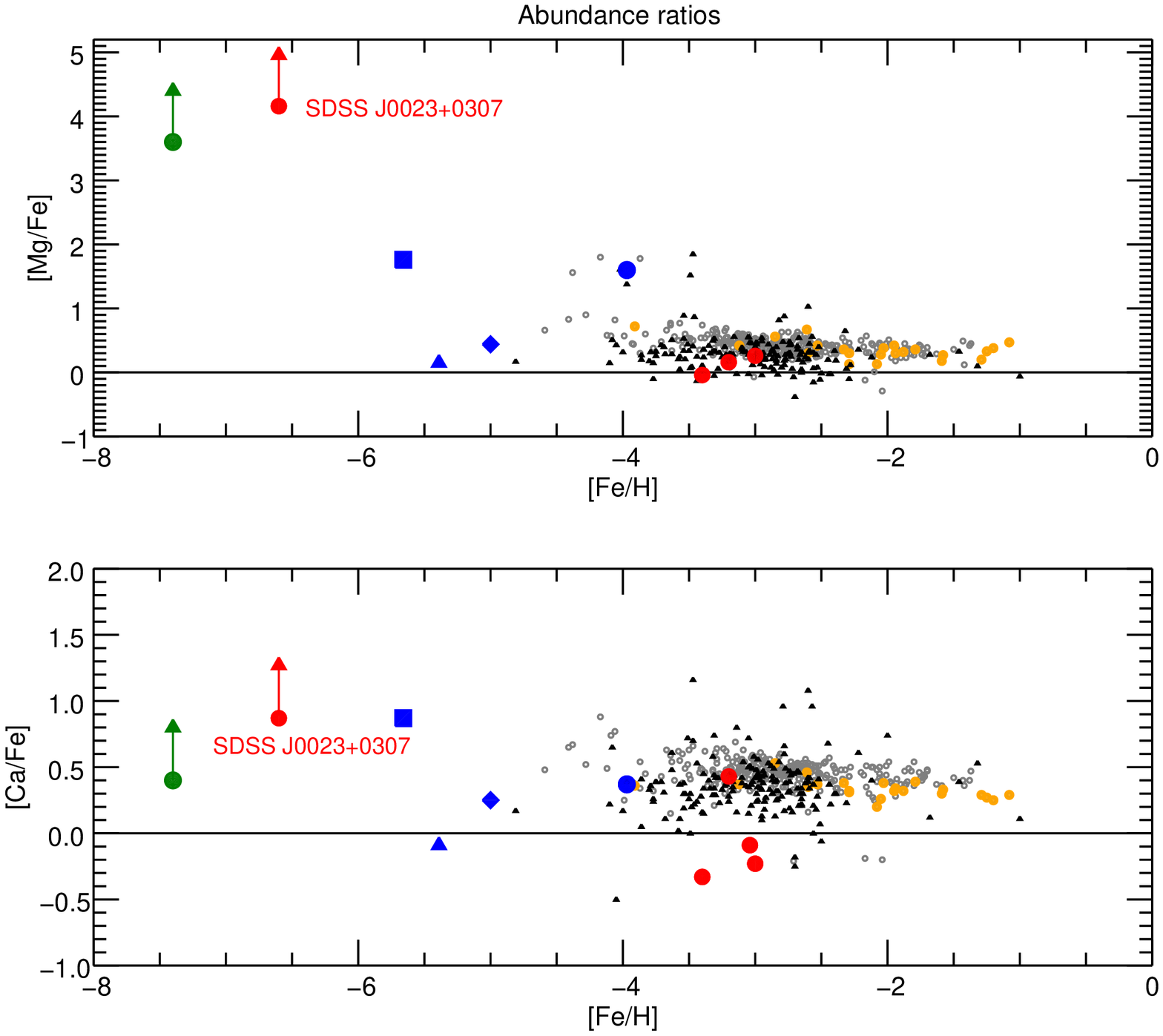}
      \caption{[Mg/Fe] and [Ca/Fe] vs [Fe/H]. Red circles : this paper; Blue circle: BPS CS22949-037 \citet{depagne2002},  green circle:  SMSS J031300$-$670839.3 \citep{keller2014};
      blue diamond:  SDSS~J1313-0019 \citep{frebel2015}; blue square:  HE1327-2329 \citep{aoki2006}; blue triangle:  HE 0107-5240 \citep{christlieb2004}; small grey circle:  evolved stars from  Roederer et al. (2014); small  orange circles: main sequence stars from Roederer (2014); black triangles: data from \citet{Yong2013}.}
         \label{alpha}
   \end{figure*}

In Figure \ref{alpha}, we plotted the abundance ratios [Mg/Fe] and [Ca/Fe] as a function of [Fe/H] for our sample of stars using the  solar abundances from \citet{lodders09}. 

It is interesting to note that three stars  of the sample are calcium poor.  As their [Mg/Fe]  abundance ratios appear also rather low, it is very likely that the origin of the low [Ca/Fe] comes 
from the [Fe/H] abundance determination, which is difficult on these  medium resolution  and rather low S/N spectra. 
 
 We adopted the [Fe/H] from \citet{aguado2018} for SDSS~J002314.00+030758.0.    As their measurements are based on spectra with  much higher
 S/N ratios ($\ge$ 170 per pixel at 450nm) than our spectra,  although of  lower resolution, 
the upper limit 
[Fe/H]
 $\le$ -4.0 dex we obtained is likely to be much higher than the true
 [Fe/H] that this star may have. Assuming an
 [$\alpha$/Fe] overabundance of 0.4 dex, a value generally found in halo stars, we derive an iron abundance of [Fe/H] =-6.13 dex.  We will consider during the discussion the possibility that this star may have a much higher iron abundance, with  [Fe/H] = -5 dex, a factor of 40 larger than the upper limit derived by  \citet{aguado2018}.
 We have added in this plot the results for evolved and unevolved stars from \citet{roederer2014} and the results from \citet{Yong2013}.
 We also include the stars CS 22949-037 \citep{depagne2002}, HE 0107-5240 \citep{christlieb2004},  HE1327-2329 \citep{aoki2006},   SDSS~J1313-0019 \citep{frebel2015}, and SMSS J031300-670839.3 \citep{keller2014}.  We could measure the magnesium abundance in four stars of our sample.  The calcium was detected and measured in five stars. 
We can see that the star SDSS~J002314.00+030758.0 stands out, revealing very high overabundances of Mg and Ca sharing this peculiarity with Keller's star (SMSS J031300-670839.3)   and to a lesser extent Depagne's star (CS 22949-037).  We could also measure  the abundance of Si  in  SDSS~J002314.00+030758.0  and found A(Si) = 4.2 dex, leading to  [Si/Fe] = 3.25 dex for an iron abundance  [Fe/H] = -6.6 dex and   [Si/Fe] = 2.75 dex for an iron abundance  [Fe/H] = -6.1 dex. In their star,  \citet{keller2014} derived an upper limit of A(Si) $\le$ 4.3 dex.
The other stars of our sample have a moderate to solar [Mg/Fe] ratio.  In SDSS~J232854.99+023924.5, we could measure the barium abundance.  With  [Ba/Fe] = 1.58 dex, this star could be classified  as a CEMP-s star.
 Unfortunately, strontium could not be measured. We could only derive an upper limit of [Sr/Fe] $\le$ 0.86 dex.

 \begin{table*}
 \caption{Fe abundances. } 
\label{Fe_abundance}
\centering
\begin{tabular}{l c c c c c c c c c c c c}
\hline\hline
  Object                            & Fe          &  Fe         & Fe        &    Fe        &   Fe         &  Fe  & Fe    & Fe &  Fe &   Fe   & Fe &  A(Fe) \\    
                       wavelength (nm)        & 356  &  357   & 378  &   382  &   383 &  386   & 388   & 392  &  438  & 440  & 527  &  \\
 
\hline 
SDSS~J232854.99+023924.5        &       ---- &  ----   &   ----       &          4.40         &    ----     &    ----       &    ----        &   ----          &      ----            &       4.40    &  4.65    & 4.50 \\
SDSS~J113207.12-082657.3        &       ---- &  ----   &   ----       &       4.46        &    ----      &      ----      &    ----           &  ----            &            ----           &        ----            & ---- & 4.46 \\
SDSS~J144551.13-004541.3        &       ---- &  ----      &  4.57     &       4.06         &   ----       &   4.34       &    4.22           & 4.72         &   4.17           &    4.21    &  -----  &  4.30 \\  
SDSS~J003730.31+245750.6        &       ---- &  ----    &    ----     &       ----          &    ----       &      ----     &            ----        &    ----     &      ----            &     ----- &  -----    &  -----   \\
SDSS~J162049.87+322323.2        &   4.65  &  4.10 &     3.60    &       ----     &    ----       &      ----     &           ----       &            ----      &   ----           &       -----    &  -----  &   4.10   \\  

  \hline
  \end{tabular}
   \end{table*}
 
\begin{table*}
 \caption{ Abundances of $\alpha$  and  neutron-capture elements . } 
\label{abundances}
\centering
\begin{tabular}{l c c c c c c c  }
\hline\hline
  Object                             &     A(C)      &    A(Mg)      &     A(Ca)      &     A(Si)      &     A(Sr)           &   A(Sr)  &     A(Ba)                    \\
      Absorption feature      &  G  band   &     b triplet   &    K line       &  390.5 nm  &     407.7 nm   &    421.5 nm  & 455.54 nm \\
\hline 
SDSS~J232854.99+023924.5        &     6.6             &    4.80         &     3.10         &   ----            &    $\le$   0.80         &  $\le$   0.80      &   0.80                        \\  
SDSS~J113207.12-082657.3        &    $\le$  6.9    &      ----         &      3.20         &    ----            &    ----              &     ----         &  ----                           \\  
SDSS~J144551.13-004541.3        &     6.4             &    4.50       &     3.50        &    ----           &   ----       &   ----       &    ----  \\  
SDSS~J003730.31+245750.6        &  $\le$   6.6       &    ----          &       ----         &     ----           &     ----            &   ----           &   ----                           \\  
SDSS~J162049.87+322323.2  &      6.4             &    4.10         &      2.60        &    3.37            &     ----            &    ----          &   ----                           \\  

  \hline
  \end{tabular}
   \end{table*}

 \section{ SDSS~J002314.00+030758.0}
 
 The abundance results  for  the star SDSS~J002314.00+030758.0 are listed in Table \ref{SDSSJ0023_abundance}. The
 assumed gravity is in agreement with the one adopted by \citet{aguado2018}.
 
  \begin{table}
 \caption{Abundances measured in  SDSS~J002314.00$+$030758.0} 
\label{SDSSJ0023_abundance}
\centering
\begin{tabular}{l l c }
\hline\hline
Element & Absorption feature &  Abundance \\
\hline\hline
Fe  &       382~nm            &         $\le$ 3.5      \\
C    &   G~band    &    6.4   \\
Mg  &   b~triplet    &    5.1   \\
Ca &    K line         &   0.6  \\
Si   & 390.5~nm    &  4.2  \\
Sr  & 407.7~nm     &     $\le$ 0.0   \\
Sr & 421.5~nm      &   ----     \\
Ba & 455.4~nm &    -----    \\

  \hline
  \end{tabular}
   \end{table}

       \begin{figure*}
   \centering
   \includegraphics[width=15cm]{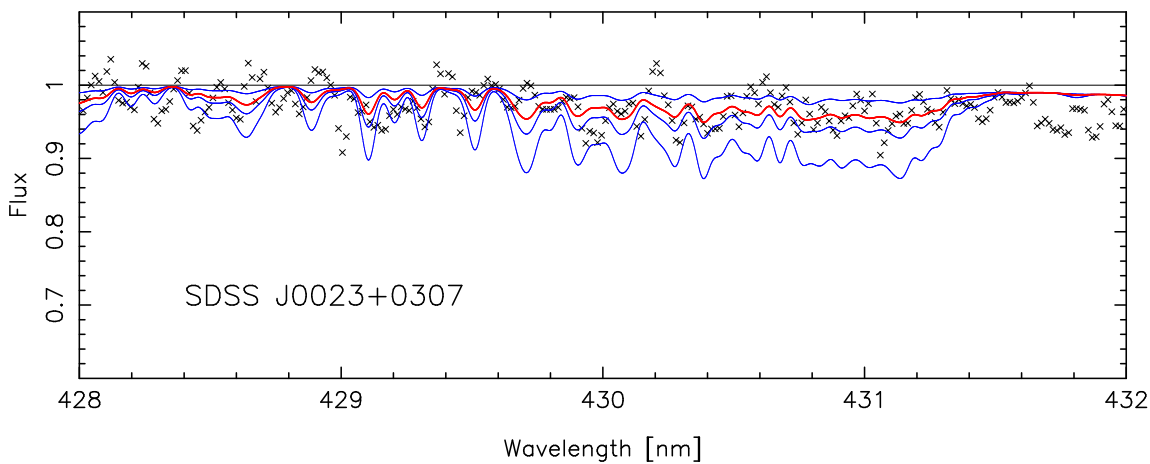}
      \caption{Synthetic spectra of SDSS~J002314.00+030758.0 in the region of the CH G-band (continuous lines) superimposed on the observed spectrum (small crosses).  The theoretical spectra are represented as blue lines correspond to carbon abundances A(C)=6.0,  6.6 and 6.9 dex. The red line represent the best fit to the observed spectrum and corresponds to a carbon abundance A(C) = 6.4 dex.}
         \label{carbon_fit}
   \end{figure*}
 
\begin{figure*}
   \centering
   \includegraphics[width=15cm]{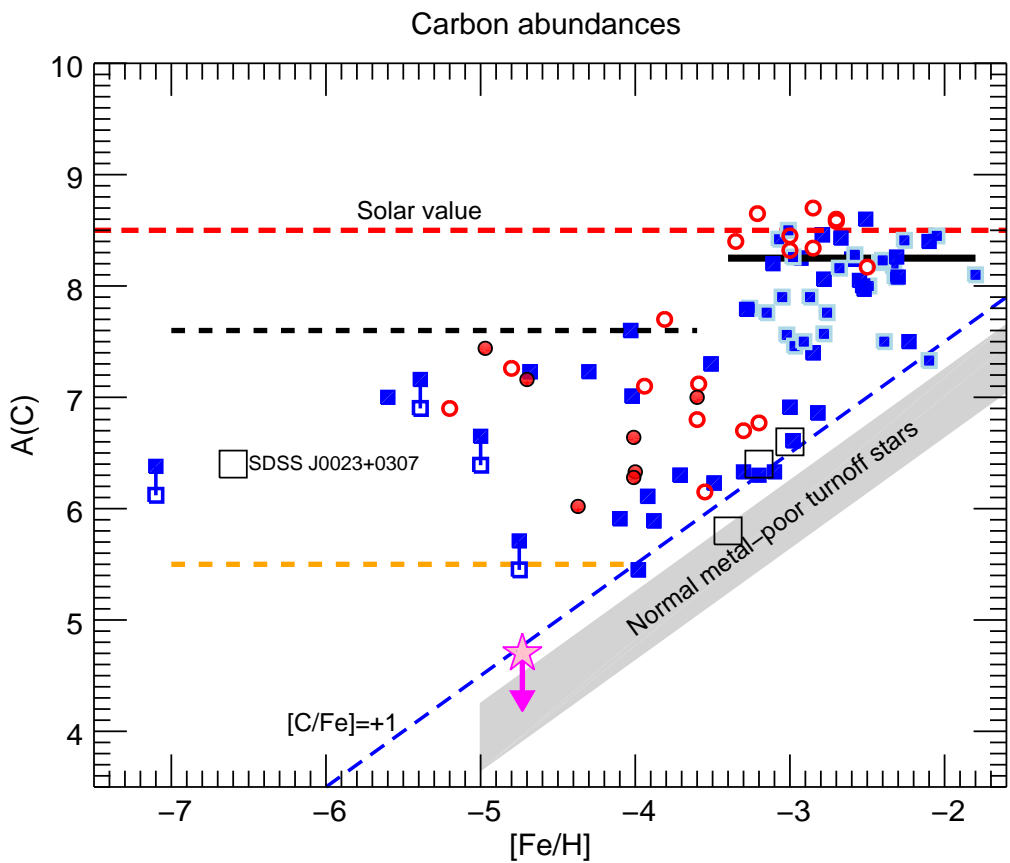}
      \caption{Carbon abundances A(C) of CEMP stars as a function of [Fe/H]. Large open black squares represent our results. The other symbols are literature data. Details can be found in  \citet{bonifacio2015}. }
         \label{carbon}
   \end{figure*}

In Fig. \ref{carbon_fit}, we plotted the computed spectrum in the region of the G-band carbon molecular band with different assumptions on the carbon abundance for the star. The best fit is found for A(C) = 6.4 dex. This result is in good agreement with the upper limit  A(C) = 6.3 dex found by \citet{aguado2018}. In Figure \ref{carbon}, we show  our stars  in the A(C)-[Fe/H] 
diagram \citep{spite2013},  together with data from the literature. When we apply the three-dimensional (3D) correction from \citet{gallagher2016} to this value, we find that it is further reduced by -0.45 dex so that A(C) = 6.05 dex.  It is interesting to note the star SDSS~J002314.00+030758.0 is located in  a region 
with CEMP stars that have a rather low carbon abundance of about A(C) = 6.5 dex. 
The bimodal distribution of CEMP stars was  first suggested by \citet{spite2013} and later supported by the extended study of
 \citet{bonifacio2015} who interpreted the low-carbon band stars as being the genuine fossil records of a gas cloud that has been enriched 
 by a faint supernova providing carbon and the lighter elements. 
 Alternatively, \citet{yoon2016} proposed a separation among the CEMP  stars  along three regions instead of the two bands system from \citet{spite2013}.

 The difference between these two classifications comes from the definition of the C-rich stars: [C/Fe]=+0.7 in \citet{yoon2016} compared to [C/Fe]=+1.0 in the present paper. 
 Our choice of the definition of the C-rich stars is based on the fact that, in the group of normal (not C-rich) stars, [C/Fe]=+0.45 $\pm$ 0.15 dex \citep{bonifacio2009}. The normal metal-poor stars are already C-rich as they are generally alpha-rich. We thus suspect that the group II of \citet{yoon2016} is only the tail of the distribution of the normal metal-poor stars.
In this classification system,  SDSS~J002314.00+030758.0  would belong to their group III of CEMP stars.  
 
 SDSS~J002314.00+030758.0  also shares the same abundance anomalies as the more metal-rich  star CS 22949$-$037 \citep{depagne2002}.  This star is a CEMP-no star with [Fe/H] $\simeq$ -4.00 dex,  [Mg/Fe]=+1.58 dex,   and [Ca/Fe] = +0.35 dex, leading to [Mg/Ca] = 1.23 dex, a remarkable overabundance shared with 
 SDSS~J002314.00$+$030758.0 and SMSS J031300$-$670839.3, in which  we found a large [Mg/Ca] ratio close to +3.2 dex. Adopting [Fe/H] = -6.6 dex, we found [Mg/Fe]  =4.16 dex and [Ca/Fe] = 0.87 dex.
  With [Fe/H] = -5.0 dex, we obtain [Mg/Fe]  =2.56 dex and [Ca/Fe] = -0.73 dex .  Such a low  [Ca/Fe] has never been found in a star with [Fe/H] $\le$ -4.0 dex.  If this star has indeed   [Fe/H] $\ge$ -5.0 dex, its chemical composition 
  would be unique. On the other hand, if we assume a    [Ca/Fe] representative of the halo population  ([Ca/Fe] = +0.4 dex), 
   we find [Mg/Fe] = +3.69 dex close to the value found by \citet{keller2014} for SMSS J031300-670839.3.
  New high quality spectra will be necessary to better constrain the [Fe/H] and  refine the value found by \citet{aguado2018}.
 
 A similar  high [Mg/Fe] ratio is also found by \citet{keller2014}  who measured a  [Mg/Ca] ratio of +2.9 dex in SMSS J031300-670839.3, and  by \citet{aoki2006}  who measured a [Mg/Ca] ratio of   +0.90 dex in  HE1327-2326.  These high [Mg/Ca] ratio stars are very peculiar and rare and so far limited to the most metal-poor stars, although this behaviour is not shared by all the stars with extremely  low metallicity. Indeed, HE~0107-5240 and SDSS~J1313-0019 have [Fe/H] below -5.0 dex and  a  [Mg/Ca] ratio corresponding closely  to a  solar  ratio as measured in the vast majority of  the stars  in the metallicity range  $\simeq$ -2.0 to $\simeq$ -4.0 dex. 
 
We gather in  Table \ref{ratios} the abundance ratios found for our sample. For SDSS~J002314.00$+$030758.0, we computed the abundance ratios with  two assumptions for the iron abundance, the first with   [Fe/H] = -6.6 dex 
 assuming the upper limit from \citet{aguado2018}     and  the second with [Fe/H] = -6.13  with [Fe/H] deduced from the calcium abundance  and assuming [Ca/Fe] = +0.4 dex. 
 
 \begin{table*}
 \caption{ [X/Fe] abundance ratios  for $\alpha$  and  neutron-capture elements. For  SDSS~J002314.00$+$030758.0, we computed the abundance ratios with [Fe/H] = -6.6 dex (a) 
 assuming the upper limit from \citet{aguado2018}  
  and [Fe/H] = -6.13 (b) with [Fe/H] deduced    from the calcium abundance  and assuming [Ca/Fe] = +0.4 dex. } 
\label{ratios}
\centering
\begin{tabular}{l c c c c c c c  }
\hline\hline
  Object                                             &     [Fe/H]      &    [C/Fe]            &     [Mg/Fe]      &   [Ca/Fe]   & [Si/Fe]      &     [Sr/Fe]          &   [Ba/Fe]                    \\
 \hline 
SDSS~J232854.99$+$023924.5      &     -3.02      &           1.12        &     0.28        &   -0.21            &             &  $\le$   0.90      &   1.65                       \\  
SDSS~J113207.12$-$082657.3      &    -3.06       & $\le$  1.46        &       ----         &   -0.07            &    ----     &     ----         &  ----                           \\  
SDSS~J144551.13$-$004541.3      &    -3.22      &            1.12               &      0.18        &    +0.39         &   ----       &   ----       &    ----  \\  
SDSS~J003730.31$+$245750.6      &     ----        &    ----                   &       ----         &     ----           &     ----       &   ----           &   ----                           \\  
SDSS~J162049.87$+$322323.2       &     -3.42     &             1.32        &     -0.02       &    -0.31          &    -0.73     &    ----          &   ----                           \\  
SDSS~J002314.00$+$030758.0  (a) &     -6.6       &             4.50       &       4.16      &      0.87         &    3.28      &    $\le$ 3.68 & ---- \\  
SDSS~J002314.00$+$030758.0  (b) &    -6.13       &            4.03       &       3.69     &     0.40           &    2.81      &   $\le$  3.21 &  ---- \\
 \hline
  \end{tabular}
   \end{table*}

In order to explain the chemical composition anomalies found in these CEMP-no metal-poor stars with a high enrichment in $\alpha$-elements, several astrophysical processes have been invoked. Faint supernovae models have been developed \citep{umeda2003, umeda2005} more recently by \citet{nomoto2013} and \citet{tominaga2014} to explain how a supernova, without sufficient explosion energy to create the heavy elements, releases 
ejecta enriched only in C, N, O, and  light metals like Na, Al , Mg, and Ca. Alternatively, \citet{hirschi2007} and \citet{meynet2010} suggested that  rapidly rotating stars,  known as spinstars, 
could produce abundance distributions similar to those found in the CEMP-no stars with a high level of enrichment  in light metals.  The wide range in [Mg/Ca] ratio 
found in the most metal-poor stars favours the existence of a diversity of nucleosynthesis conditions where magnesium can be produced in much larger quantities than calcium.  In conclusion, this is consistent with an environment of formation for CEMP-no stars polluted only 
by zero-metallicity stars \citep[e.g.][]{debennassuti2017,  bonifacio2015}, which  most likely had different masses and possibly different explosion energies that could produce various amount of Mg. 

\section{Conclusions}

In this article, we reported  the chemical analysis of six new extremely  metal-poor candidates observed with the X-Shooter spectrograph. 
We  determined the abundances of some elements (C, Mg, Ca, Si, Sr, and Ba) in the majority of these stars.  Five stars of the sample 
show abundance ratios that are typical of metal-poor stars in the metallicity range  -3.5 dex $\le$  [Fe/H] $\le$ -2.0 dex. 
The dwarf star SDSS~J002314.00+030758.0 appears to be extremely  iron-poor. We found an upper limit of [Fe/H] $<$ -4.00 dex, a value which is in agreement with the even stronger upper limit computed by \citet{aguado2018} with  [Fe/H] $<$ -6.60 dex.  Our low S/N ratio does not allow us to give a more constraining [Fe/H] determination. Assuming a    [Ca/Fe] representative of the halo population  ([Ca/Fe] = +0.4 dex), we obtain [Fe/H] = -6.1 dex.
We could determine the carbon abundance using the G band. We obtained A(C) = 6.4 dex (6.05 in 3D), which places this star in the lower band of 
the A(C)-[Fe/H] diagram \citep{spite2013, bonifacio2015} confirming the existence of the lower carbon band in the most metal-poor stars. 
Assuming the lower limit [Fe/H] = -6.6 dex from \citet{aguado2018} as a conservative estimate of its iron content,  the [C/Fe] ratio would give  $\simeq$ +4.6 dex. Assuming [Fe/H] =-5 dex, 
the [C/Fe] remains extremely high with a value of  [C/Fe] = +3.0 dex.
Given the large amount of carbon present in this star, its total metallicity lies in the range of  the metallicities found in Galactic  globular clusters.

Adopting the same lower limit [Fe/H] = -6.6 dex from \citet{aguado2018}, we found that  SDSS~J0023+0307 has  remarkably high magnesium and calcium abundances,  sharing this peculiarity with CS 22949-037 \citep{depagne2002}, SMSS J031300-670839.3 \citep{keller2014}, and HE~1327-2326 \citep{aoki2006}. These four stars have also a high [Mg/Ca] ratio ($>$ 0.90 dex) in contrast with the other extremely iron-poor stars HE~0107-5240  \citep{christlieb2004} and SDSS~J1313-0019  \citep{frebel2015}, suggesting different channels for the enrichment of the gas that formed most metal-poor stars we observe today.

\begin{acknowledgements}
This work was supported by JSPS and CNRS under the Japan-France Research Cooperative Program (CNRS PRC No 1363), the JSPS Grants-in-Aid for Scientific Research (26400232, 26400237), and the RIKEN iTHEMS Project.
P.F. acknowledges support by the Conseil Scientifique de l'Observatoire de Paris.
S.S. acknowledges support by the Italian Ministry of Education, University, 
and Research (MIUR) through a Rita Levi Montalcini Fellowship.
AJG acknowledges the Sonderforschungsbereich SFB 881 "The Milky Way System" (subproject A5) of the German Research Foundation (DFG).
 In memoriam to  our dear colleague Yuhri Ishimaru (1967-2017), who was member of this collaboration.
\end{acknowledgements}

%
%

\end{document}